\documentclass[12pt,preprint]{aastex}

\slugcomment{This manuscript is accepted by \textbf{Ap\&SS}}

\shorttitle{Magnetic fields and matter infall}
\shortauthors{Nejad-Asghar}

\begin{document}

\title{Matter Infall in Collapsing Molecular Cloud Cores with an Axial Magnetic Field}

\author{Mohsen Nejad-Asghar}

\affil{Department of Physics, University of Mazandaran, Babolsar,
Iran}

\email{nejadasghar@umz.ac.ir}

\begin{abstract}
The magnetic fields affect collapse of molecular cloud cores. Here,
we consider a collapsing core with an axial magnetic field and
investigate its effect on infall of matter and formation of
accretion disk. For this purpose, the equations of motion of ions
and neutral infalling particles are numerically solved to obtain the
streamlines of trajectories. The results show that in non-steady
state of ionization and ion-neutral coupling, which is not
unexpected in the case of infall, the radius of accretion disk will
be larger as a consequence of axial magnetic field.
\end{abstract}


\section{Introduction}

A lot of observational information is now known about the structure
of dense cores as progenitors of stars within the molecular clouds
(e.g., Evans~2010). Theoretically, we expect that most of these
cores will collapse into protostars, but the details of evolutions
are less evident. Clearly, the evolution heavily depends upon the
effects of local density, pressure, rotation, magnetic fields, the
presence or absence of nearby stars and protostars, and upon the
other physical phenomena. The most commonly used picture of low-mass
star formation is inside-out collapse (Shu~1977) in which it begins
at the center of a singular isothermal sphere and an expanding wave
of infalling matter propagates outward with the speed of sound. Up
to now, different models for core collapse and formation of
protostars have been presented, in each the effects of some physical
phenomena investigated (e.g., McKee and Ostriker~2007).

The velocity maps of molecular cloud cores, which are built by using
the velocity measurements along the line of sight, show a small
velocity gradient across each core. Measuring this global velocity
gradient, with assumption that core has a uniform rotation and
follow a rigid-body rotation law, can be used to deduce the core
angular velocity (e.g., Goodman et al.~1993). The large size of
initial collapsing core implies that even the modest initial
rotational velocities will cause the infalling matters to land first
on a rotationally supported disk rather than a pressure-supported
proto-star (e.g., Hartmann~2009, Nejad-Asghar~2011). In the simplest
analysis of rotating collapse, we assume that pressure forces are
negligible and so the problem can be analyzed by using ballistic
trajectories. The results for collapse of a spherically symmetric
cloud in uniform (solid-body) rotation were initially worked out by
Ulrich~(1976), with subsequent extension to disk formation by Cassen
and Moosman~(1981), and to collapsing singular isothermal sphere by
Terebey, Shu and Cassen~(1984). Mendoza, Tejeda and Nagel~(2009,
hereafter MTN) have recently generalized this idea by construction a
steady analytic accretion flow model for a finite rotating gas
cloud. They show that the streamlines and density profiles deviate
considerably from ones calculated by Ulrich, and for all relevant
astrophysical situations, the assumption of a finite cloud radius
needs to be used.

The observations indicate that magnetic energy in the molecular
clouds is comparable to the gravitational energy (e.g.,
Crutcher~1999). Moreover, the magnetic fields are also theoretically
believed to play an important role in gravitational collapse of the
molecular cloud cores. They provide pressure support against the
gravity and carries away angular momentum prior to and during the
collapse of cores to form accretion disks, jets and protostars
(e.g., Machida~2010). Although, in a molecular cloud core, the
spatial configuration of magnetic field lines is not simple (e.g.,
Whittet~2005), but polarimetry observations of young stellar objects
suggest that circumstellar thin disks around the young stars have
approximately aligned perpendicular to the magnetic fields (e.g.,
Pereyra et al.~2009). Here, we consider a simple initial
configuration in which the magnetic fields are assumed to be
parallel with rotational axis of core. Since the molecular cloud
cores are lightly-ionized, the ambipolar diffusion in which the
magnetic field is frozen into the charged species and drifts along
with them through neutrals, is an important mechanism (e.g.,
Adams~2009). Thus, we consider the effect of magnetic fields
directly on charged particles, while the neutral species feel them
indirectly via the collisions with ions.

In this way, as an extension to the work of MTN, we investigate the
effect of axial magnetic field on streamlines of infalling ions and
neutral particles. For this purpose, formulation of problem and
equations of motion are given in section~2. In section~3,
trajectories of infalling ions and neutral particles are
investigated and effect of the magnetic fields on the accretion disk
radius is studied. Finally, section~4 is allocated to summary and
conclusions.

\section{Formulation of problem}

The molecular cloud cores are lightly-ionized gases with small
fraction of ions with density $n_i$ and electrons with density $n_e
\approx n_i$. If the volumetric rate of electron-ions recombination
took place in gas phase, it would be proportional to $n_e n_i
\propto n_i^2$. In a simple steady state approximation, the
volumetric rate of electron-ions recombination $\propto n_i^2$ would
be equal to the volumetric rate of neutrals ionization via cosmic
rays $\propto n_n$. Although, the actual ionization balance in the
molecular clouds is more complicated, Elmegreen~(1979) and
Umebayashi and Nakano~(1980) showed that the relation $n_i \propto
n_n^{1/2}$ is a good approximation. Here, we suppose for fiducial
purpose that $\rho_i = \tilde{\epsilon} \epsilon \rho_n^{1/2}$,
where $\epsilon = 9.5 \times 10^{-15} \mathrm{m^{-3/2}kg^{1/2}}$ and
$\tilde{\epsilon}$ is a dimensionless free parameter which
represents the deviations from calculations in the steady state
approximation.

In the magnetized cloud cores, the mean velocity $\textbf{v}_n$ of
neutrals will not generally equal to the mean velocity
$\textbf{v}_i$ of ions and $\textbf{v}_e$ of electrons. Although,
the positive ions and electrons feel the magnetic forces in opposite
direction, but the induced electric field will generally cause they
move in ambipolar motion, so that in the time-scales that are
considered here, their mean velocities are approximately the same
($\textbf{v}_e \approx \textbf{v}_i$). Since electrons carry much
less momentum than ions, we neglect the dynamic of electrons. In
this way, resistance of relative drift between ions and neutrals
will be a drag acceleration arises from mutual collisions between
them. The drag acceleration exerted on neutrals by ions is
$\tilde{\gamma} \gamma \rho_i (\textbf{v}_i - \textbf{v}_n)$; the
drag acceleration exerted on ions by neutrals is $- \tilde{\gamma}
\gamma \rho_n (\textbf{v}_i - \textbf{v}_n)$, where $\gamma=3.5
\times 10^{10} \mathrm{m^3 kg ^{-1} s^{-1}}$ (Draine, Roberge and
Dalgarno~1983) is the drag coefficient and $\tilde{\gamma}$ is a
dimensionless free parameter which represents the deviations from
calculations in the steady state approximation.

The equations of motion for ions and neutral particles, in
gravitational field of a central mass, are respectively as follows
\begin{equation}\label{equmotioni}
    \ddot{\textbf{r}}_i = -\frac{GM}{r_i^3} \textbf{r}_i + \frac{q_i}{m_i}
    \dot{\textbf{r}}_i \times \textbf{B} - \tilde{\gamma} \gamma \rho_n
    (\dot{\textbf{r}}_i - \dot{\textbf{r}}_n),
\end{equation}
\begin{equation}\label{equmotionn}
    \ddot{\textbf{r}}_n = -\frac{GM}{r_n^3} \textbf{r}_n +
    \tilde{\epsilon} \tilde{\gamma} \epsilon \gamma \rho_n^{1/2} (\dot{\textbf{r}}_i - \dot{\textbf{r}}_n),
\end{equation}
where $M$ is the proto-stellar mass at the origin, and $q_i=N_1e$
and $m_i=N_2m_p$ are ion charge and mass, respectively, as an
integer multiple of the electron charge $e$ and the proton mass
$m_p$. Here, we assume that $N_2 \approx 2 N_1$.

We measure the length and time in units of $0.01 \mathrm{pc}$ and
$10^5 \mathrm{yr}$, respectively, so that velocity unit is
$0.1~\mathrm{km\;s^{-1}}$. Mass unit is appointed to $1 M_\odot$,
thus the gravitational constant is $G= 45$. In this manner, we have
$\gamma=7.8\times 10^9$ and $\epsilon=3.5 \times 10^{-8}$. Finally,
the unit of magnetic field is chosen equal to
$10~\mathrm{nT}=100~\mu G$. Using these units, the equations of
motion (\ref{equmotioni}) and (\ref{equmotionn}) can be rewritten in
a dimensionless form as follows
\begin{equation}\label{equmononi}
    \ddot{\textbf{r}}_i = -45 \frac{M}{r_i^3} \textbf{r}_i + 0.5 \tilde{B}
    \dot{\textbf{r}}_i \times \hat{\textbf{B}} - 3.9 \times 10^5 \tilde{\gamma}
    \tilde{n}_n (\dot{\textbf{r}}_i - \dot{\textbf{r}}_n),
\end{equation}
\begin{equation}\label{equmononn}
    \ddot{\textbf{r}}_n = -45 \frac{M}{r_n^3} \textbf{r}_n +
    1.9 \tilde{\epsilon} \tilde{\gamma} \tilde{n}_n^{1/2}
    (\dot{\textbf{r}}_i - \dot{\textbf{r}}_n),
\end{equation}
where $\hat{\textbf{B}}$ is a unit vector in the direction of
magnetic field, and $\tilde{n}_n\equiv n_n/(10^{10}
\mathrm{m^{-3}})$ and $\tilde{B}$ are two dimensionless free
parameters.

Choosing the axial magnetic field $\hat{\textbf{B}}= \hat{k}$, the
equations of motion in Cartesian coordinate are
\begin{equation}\label{xiddot}
    \ddot{x}_i =-45M\sin\theta_i \cos\phi_i \frac{1}{r_i^2} + 0.5\tilde{B}
    \dot{y}_i - 3.9 \times 10^5 \tilde{\gamma}
    \tilde{n}_n
    (\dot{x}_i - \dot{x}_n),
\end{equation}
\begin{equation}\label{yiddot}
    \ddot{y}_i =-45M\sin\theta_i \cos\phi_i \frac{1}{r_i^2} - 0.5\tilde{B}
    \dot{x}_i - 3.9 \times 10^5 \tilde{\gamma}
    \tilde{n}_n
    (\dot{y}_i - \dot{y}_n),
\end{equation}
\begin{equation}\label{ziddot}
    \ddot{z}_i = -45M\sin\theta_i \cos\phi_i \frac{1}{r_i^2} - 3.9 \times 10^5 \tilde{\gamma}
    \tilde{n}_n
    (\dot{z}_i - \dot{z}_n),
\end{equation}
\begin{equation}\label{xnddot}
    \ddot{x}_n =-45M\sin\theta_n \cos\phi_n \frac{1}{r_n^2} + 1.9 \tilde{\epsilon} \tilde{\gamma}
    \tilde{n}_n^{1/2} (\dot{x}_i - \dot{x}_n),
\end{equation}
\begin{equation}\label{ynddot}
    \ddot{y}_n =-45M\sin\theta_n \cos\phi_n \frac{1}{r_n^2} + 1.9 \tilde{\epsilon} \tilde{\gamma}
    \tilde{n}_n^{1/2} (\dot{y}_i - \dot{y}_n),
\end{equation}
\begin{equation}\label{znddot}
    \ddot{z}_n =-45M\sin\theta_n \cos\phi_n \frac{1}{r_n^2} + 1.9 \tilde{\epsilon} \tilde{\gamma}
    \tilde{n}_n^{1/2} (\dot{z}_i - \dot{z}_n),
\end{equation}
where $\theta_{i}$, $\theta_{n}$, $\phi_{i}$, and $\phi_{n}$ are the
polar and azimuthal angles of ions and neutral particles,
respectively, and $r_i=\sqrt{x_i^2+ y_i^2 + z_i^2}$ and
$r_n=\sqrt{x_n^2+ y_n^2 + z_n^2}$ are their radial distances. There
are four free parameters $\tilde{B}$, $\tilde{\epsilon}$,
$\tilde{\gamma}$, and $\tilde{n}_n$ which represent the strength of
magnetic field, ionization fraction, drag coefficient, and neutral
density, respectively.

\section{Trajectories of infalling matters}

As mentioned before, there are many observations that show velocity
gradients in the maps of molecular cloud cores. These velocity
gradients can be used to deduce the rotation of cores, but
evaluation of their angular velocity $\Omega_0$ is considerably
complicated because the velocity fields usually exhibit complex
supersonic motions. Here, we choose the fiducial value of
$\Omega_0=3\times10^{-13} \mathrm{rad\;s^{-1}}$, which in the time
unit $10^5 \mathrm{yr}$ is $\Omega_0\approx 1$. The effects of
initial angular momentum in core collapse and formation of accretion
disks are now approximately well known (e.g., Hartmann~2009). Here,
we turn our attention to the effects of magnetic fields and drag
force on infalling matters. If we assume that $\tilde{\gamma}=0$ and
$\tilde{B}=0$, equations (\ref{xnddot})-(\ref{ziddot}) lead to the
trajectories which are contained on a plane and given by a
\textit{conic section} (i.e., the results obtained by MTN).

\begin{figure}
\epsscale{.33} \center \plotone{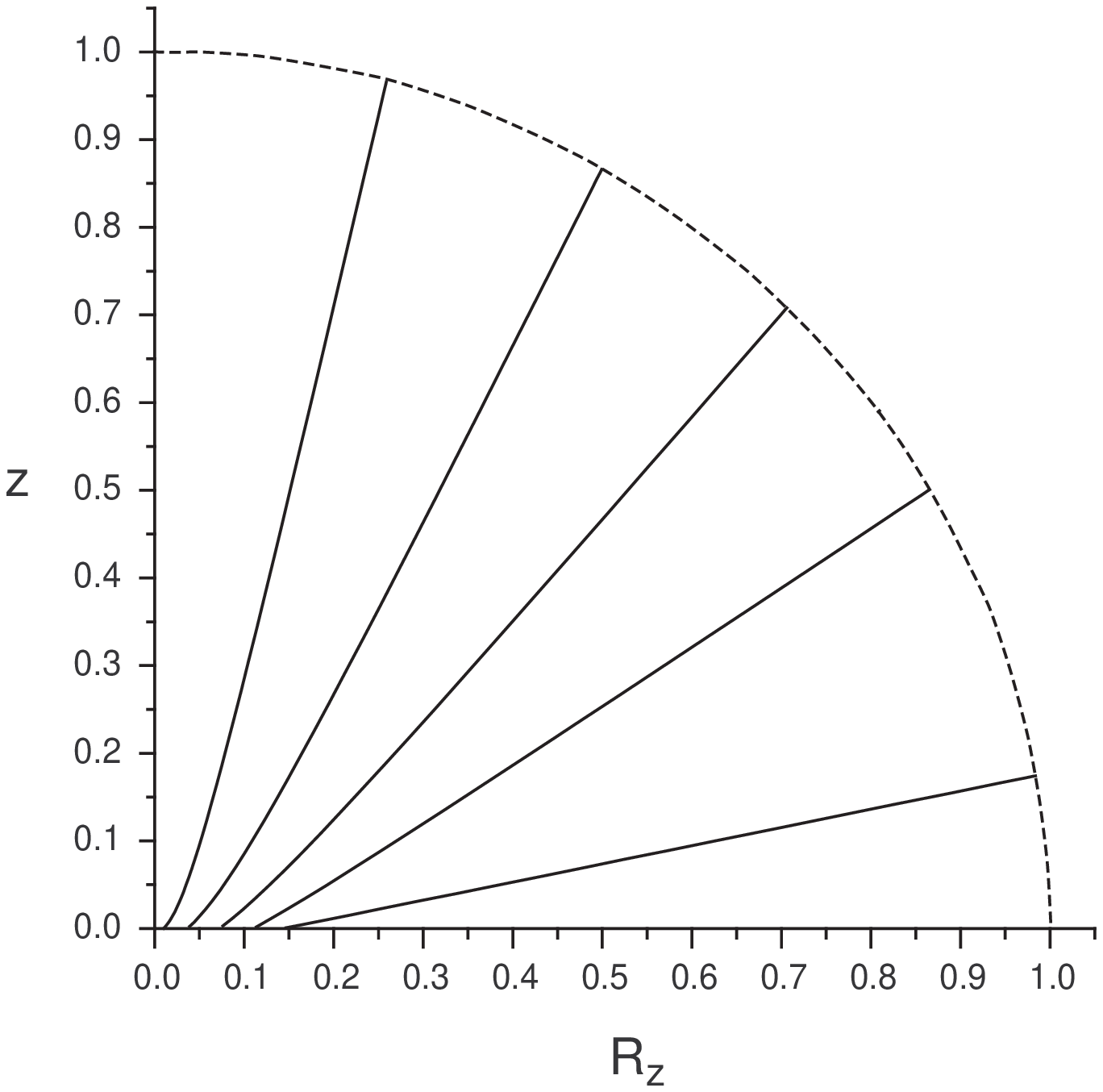}\\{(a)}\\
\epsscale{.35} \center \plotone{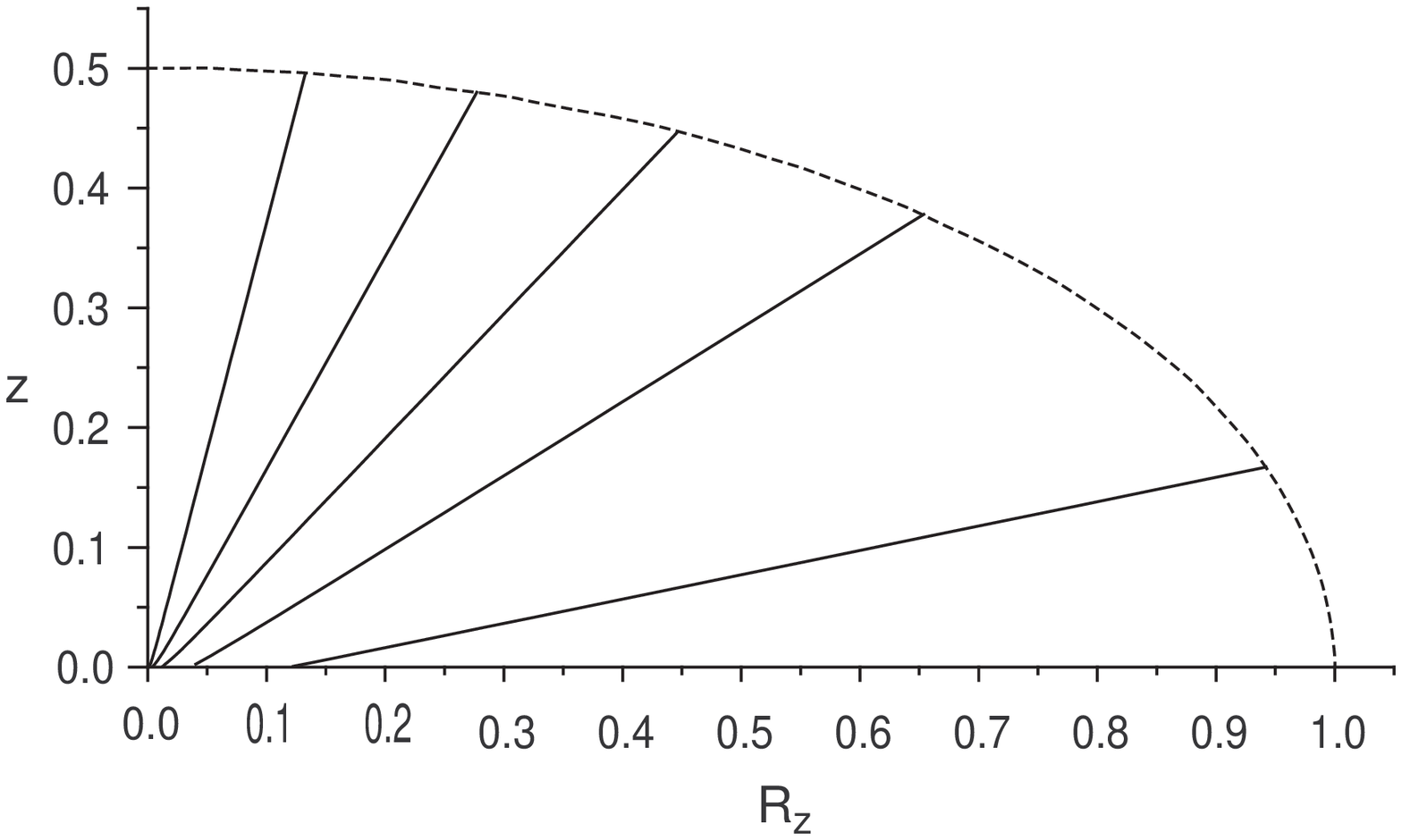}\\{(b)}\\
\epsscale{.20} \center \plotone{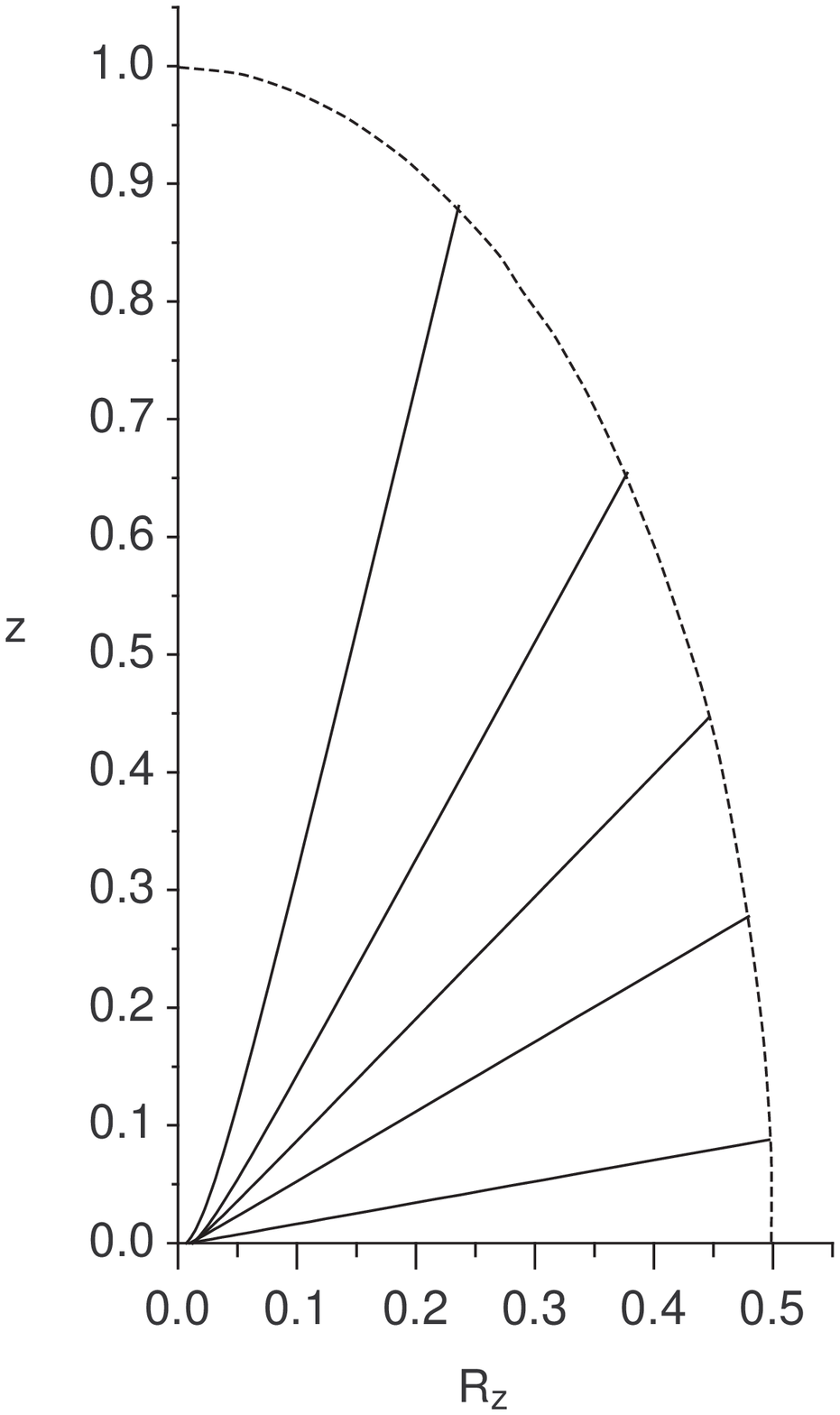}\\{(c)}\\
\caption{The streamlines of ions and neutral particles in absence of
magnetic field and initial radial velocity ($\tilde{B}=0$,
$\tilde{\gamma}=0$, $v_{0r}=0$), for (a) initial spherical core
($r_0=1$), (b) initial oblate core ($r_0= 1/\sqrt{\sin^2\theta_0 +4
\cos^2 \theta_0}$), and (c) initial prolate core ($r_0=
1/\sqrt{4\sin^2\theta_0 + \cos^2 \theta_0}$). In each figure, the
curves from top to bottom have the initial azimuthal angle equal to
$\theta_0=15$, $30$, $45$, $60$, and $80^\circ$.\label{zRso}}
\end{figure}

Choosing the initial conditions
\begin{eqnarray}\label{r0}
   \nonumber \textbf{r}_n=\textbf{r}_i&=&r_0 \hat{r}\\&=& r_0 \sin\theta_0 \cos
    \phi_0 \hat{i} + r_0 \sin\theta_0 \sin \phi_0 \hat{j} + r_0 \cos\theta_0 \hat{k},
\end{eqnarray}
\begin{eqnarray}\label{v0}
   \nonumber \textbf{v}_n=\textbf{v}_i = -v_{0r} \hat{r} + v_{0\phi}
    \hat{\phi}
    &= &-(v_{0r} \cos \phi_0 +r_0 \sin\phi_0)\sin\theta_0 \hat{i}
    \\ &&-(v_{0r}\sin\phi_0 - r_0 \cos\phi_0)\sin\theta_0
    \hat{j} -v_{0r} \cos \theta_0 \hat{k},
\end{eqnarray}
for position and velocity of infalling particles, the equations of
motion (\ref{xiddot})-(\ref{znddot}) can be solved by numerical
methods such as Runge-Kutta. Approximately, all cores in the maps of
molecular clouds seem apparently to be elongated rather than
spherical (e.g., Curtis and Richer~2010). Since determining the
exact three-dimensional shape of a core from apparent observations
of the-plane-of-sky is impossible, statistical techniques have to be
applied; as the results, some works indicate a preference for
prolate cores (e.g., Gammie et al.~2003, Li et al.~2004), and some
others favour oblate shapes (e.g., Curry~2002, Jones and Basu~2002).
Theoretically, the prolate elongation of the cores may be inferred
as a remnant of their origin in filaments and turbulent flows (e.g.,
Hartmann~2002), while the oblate shape of collapsing cores may be
justified from the effects of magnetic fields and rotational motion
(e.g., Nejad-Asghar~2010). The streamlines of infalling particles,
in absence of magnetic field ($\tilde{B}=0$, $\tilde{\gamma}=0$) and
initial radial velocity ($v_{0r}=0$), for initial spherical core
($r_0=1$), oblate core ($r_0= 1/\sqrt{\sin^2\theta_0 +4 \cos^2
\theta_0}$), and prolate core ($r_0= 1/\sqrt{4\sin^2\theta_0 +
\cos^2 \theta_0}$) are shown in Fig.\ref{zRso}.

\begin{figure}
\epsscale{1} \center \plotone{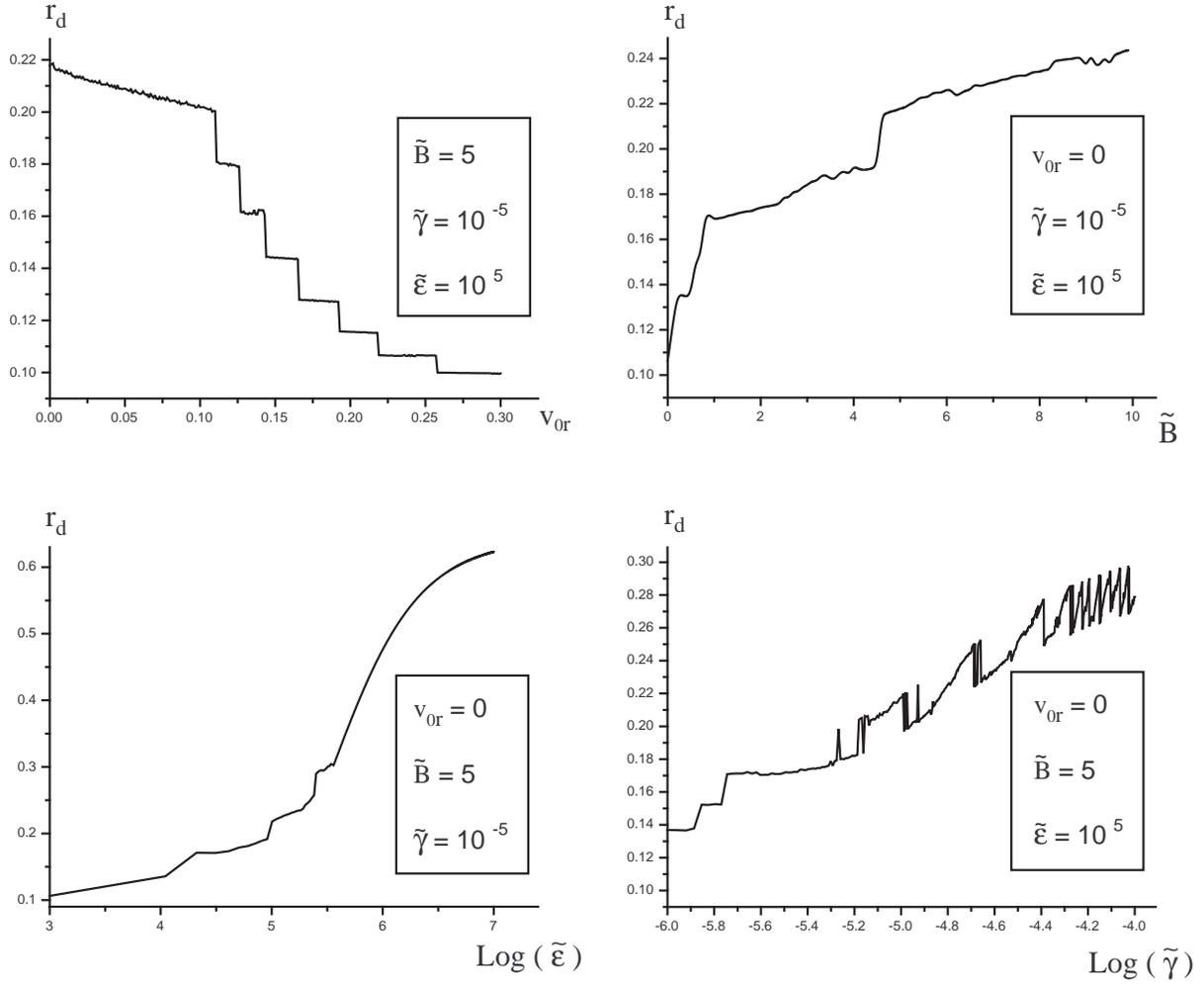} \caption{Accretion disk
radius by the substitution $\theta_0 \rightarrow \pi/2$ for various
initial free parameters with $\tilde{n}=1$.\label{rdisk}}
\end{figure}

\begin{figure}
\epsscale{.55} \center \plotone{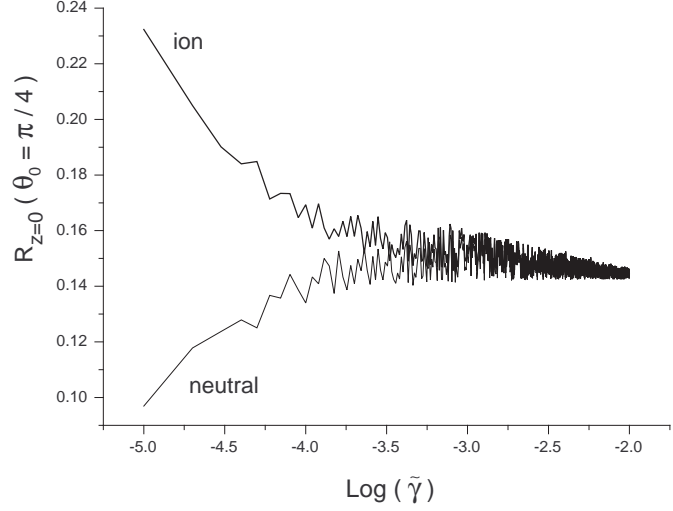}\\{(a)}\\
\epsscale{.55} \center \plotone{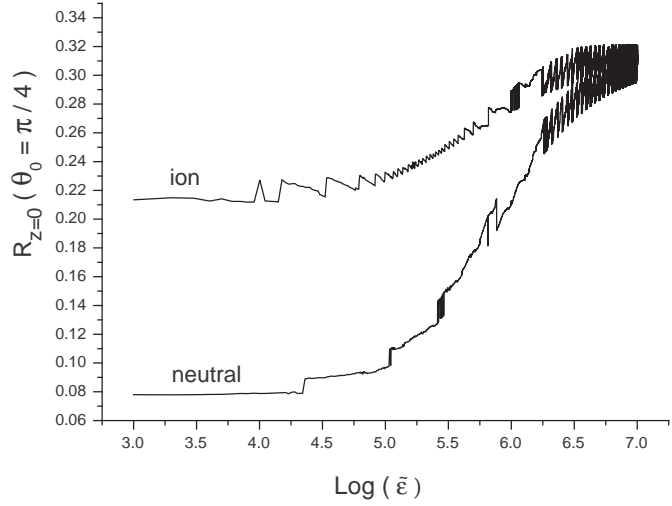}\\{(b)}\\
\caption{The equatorial radius of infalling ions and neutral
particles which start at $\theta_0=\pi/4$, versus (a) logarithm of
free parameter $\tilde{\gamma}$ with $\tilde{\epsilon}=10^5$, and
(b) logarithm of free parameter $\tilde{\epsilon}$ with
$\tilde{\gamma}=10^{-5}$. In these figures the free parameters
$\tilde{B}$ and $\tilde{n}_n$ are chosen equal to $5$ and $1$,
respectively. \label{Rtogameps}}
\end{figure}

The numerical plots of accretion disk radius for neutral particles
by substitution $\theta_0 \rightarrow \pi/2$, for a wide variety of
initial parameters, are shown in Fig.~\ref{rdisk}. The numerical
results show that the accretion disk radius is approximately
proportional to the inverse of initial radial velocity $v_{0r}$ as
was given by the analytical ballistic models of MTN. Since we want
to investigate the effect of magnetic fields on trajectories of
infalling matters, without losing the generality of problem, we
consider an initial spherical core with assumption $v_{0r}=0$. Here
we turn our attention to only one streamline with $r_0=1$ and
$\theta_0=\pi/4$. The effects of axial magnetic field on this
streamline are shown in Fig.\ref{zRdiffmag} for three different
values of free parameter $\tilde{B}$ with a fixed value of three
parameters $\tilde{\gamma}$, $\tilde{\epsilon}$ and $\tilde{n}_n$
equal to $10^{-5}$, $10^5$, and $1$, respectively. As a result, we
see that the axial magnetic field causes the infalling ions land at
a larger equatorial radius, and collisions of ions with neutral
particles lead to the same behavior for infalling neutral matters.
In Fig.\ref{Rtogameps}, the equatorial radii of ion and neutral
infalling particles (from $\theta_0=\pi/4$) are depicted versus
$\tilde{\gamma}$ and $\tilde{\epsilon}$ at a fixed value of magnetic
field. We see that increasing of drag coefficient $\tilde{\gamma}$
leads to coupling of ions and neutrals so that their equatorial
radius reach to each other asymptotically. The same behavior can be
seen for the effect of ionization degree $\tilde{\epsilon}$ so that
increasing of this parameter asymptotically causes to approximately
the same equatorial radii of ions and neutral infalling particles.
The third free parameter is neutral density $\tilde{n}_n$ which its
effect is the same as $\tilde{\gamma}$ but its effect on neutral
particles is less.

\begin{figure}
\epsscale{0.7} \center \plotone{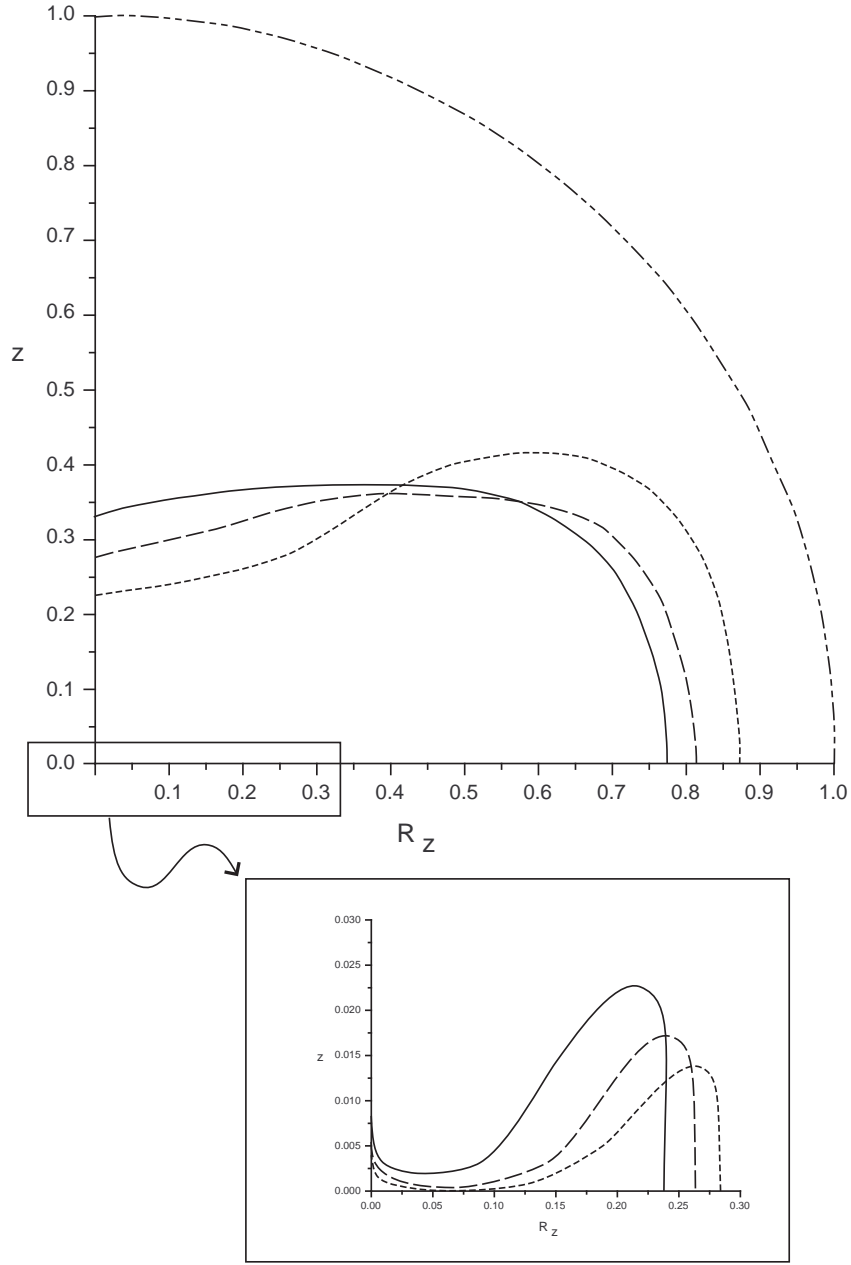} \caption{The typical
iso-density contours for $\rho\approx 0.01$ (top panel), and
$\rho\approx 0.1$ (bottom panel), for different values of free
parameter $\tilde{B}=1$ (solid), $\tilde{B}=5$ (dash), and
$\tilde{B}=10$ (dot), with $\tilde{\gamma}=10^{-5}$,
$\tilde{\epsilon}= 10^5$ and $\tilde{n}_n = 1$. \label{isoden}}
\end{figure}

The density of neutral particles can be evaluated by assuming that
the mass infall rate, $\dot{M}$, is steady. Assuming that the cloud
at $r_0$ is nearly spherical, the mass flow in a flow tube spanned
by $\triangle\theta_0$ is
\begin{equation}\label{dens1}
    \triangle \dot{M} = \frac{2\pi r^2 \sin \theta_0  \dot{M}\triangle\theta_0}
    {4 \pi r^2} = \frac{1}{2} \dot{M} \sin\theta_0 \triangle
    \theta_0.
\end{equation}
The density at $(r,\theta)$ can be found by following the
streamlines corresponding to $\theta_0$ and
$\theta_0+\triangle\theta_0$,
\begin{equation}\label{dens2}
    \rho(r,\theta) = \frac{\triangle \dot{M}}{2\pi r^2 \sin \theta \triangle
    \theta |v_r|} = \frac{\dot{M}}{4\pi r^2 |v_r|} \frac{\sin \theta_0}{\sin \theta}
    \frac{\triangle \theta_0}{\triangle \theta}.
\end{equation}
In order to obtain the iso-density contours, we divide $0 \leq
\theta_0 \leq \pi/2$ to $N$ equal segments so that $\triangle
\theta_0 = \pi/2N$. For two streamlines in the borders of each
segment, values of $\sin\theta_0$, $\sin\theta$ and $v_r$ are
evaluated in the mid points. In this way, the density at different
radii and polar angles can be found from (\ref{dens2}), and the
iso-density curves can be depicted as typically are shown in
Fig.~\ref{isoden}. In this figure, the mass accretion rate is
assumed to be $\dot{M}=4\pi\times10^{-7} \mathrm{M_\odot} \;
\mathrm{yr^{-1}}$. These iso-density contours show that the axial
magnetic fields cause the infalling matters land at a larger
equatorial radius.

\section{Summary and conclusions}

Nowadays, we have accepted that the molecular cloud cores can be
gravitationally unstable under certain conditions, and they can
collapse. Rotational motion of the cores leads to land the infalling
matters on an accretion disk around the proto-star. The streamlines
of infalling matters for initial spherical, oblate and prolate cores
are depicted in Fig.\ref{zRso}. Obviously, the centrifugal force
causes to increase the initial dimension of accretion disk in oblate
cores, while it decreases the disk radius in prolate ones.

\begin{figure}
\epsscale{.55} \center \plotone{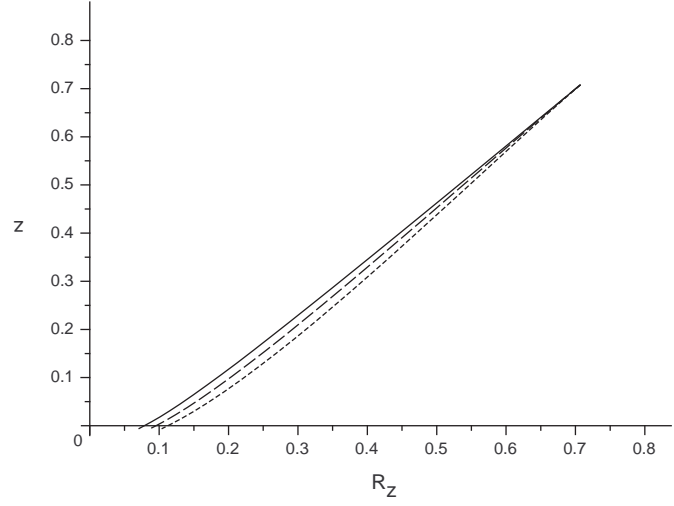}\\{(a)}\\
\epsscale{.55} \center \plotone{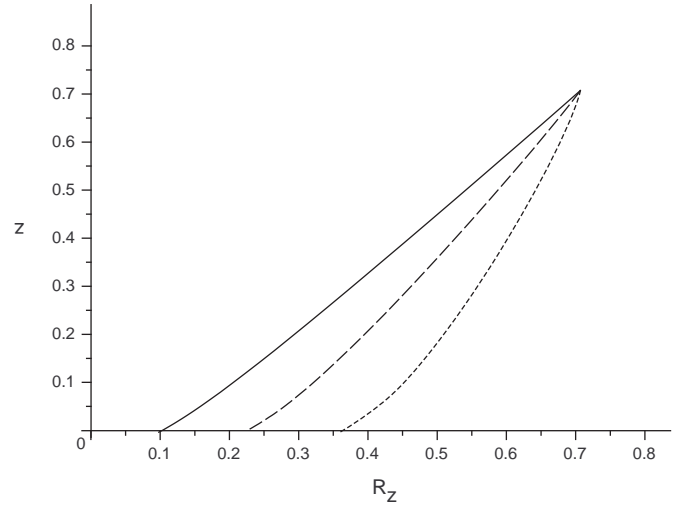}\\{(b)}\\
\caption{The streamlines of (a) neutral and (b) ion flows for
different values of free parameter $\tilde{B}=1$ (solid),
$\tilde{B}=5$ (dash), and $\tilde{B}=10$ (dot),  at $\theta_0=
\pi/4$ with $\tilde{\gamma}=10^{-5}$, $\tilde{\epsilon}= 10^5$ and
$\tilde{n}_n = 1$. \label{zRdiffmag}}
\end{figure}

Next, we considered an axil magnetic field in the collapsing core.
For investigation the effect of this magnetic field, we wrote
separately the equations of motion of neutral and ion infalling
particles. In general form, these equations are presented with four
free parameters $\tilde{B}$, $\tilde{\epsilon}$, $\tilde{\gamma}$,
and $\tilde{n}_n$ which represent the strength of magnetic field,
ionization fraction, drag coefficient, and neutral density,
respectively. The accretion disk radius for different values of the
relevant input parameters are depicted in Fig.\ref{rdisk}. Since our
goal is to study the effect of magnetic field and drag coefficient
on infalling matters and disk formation, we turned our attention to
only one streamline at $\theta_0=\pi/4$; the results are shown in
Fig.\ref{zRdiffmag}.

In a steady state of molecular gas, the ions are well-coupled to
neutral particles (i.e., $\tilde{\epsilon}=1$, $\tilde{\gamma}=1$)
so that their dynamics are approximately the same.
Fig.\ref{Rtogameps} shows that increasing of drag coefficient and
ionization degree lead to the same behavior for ions and neutral
infalling particles. In infalling case, the frequency of collisions
between ions and neutral particles eventually becomes so low that
the balance cannot be maintained, thus, the values of free
parameters $\tilde{\epsilon}$ and $\tilde{\gamma}$ deviate from the
steady sate. In this non-equilibrium case, ions are directly
affected by magnetic field so that they land in a larger radii at
equatorial plane. In the low-coupling ($\tilde{\gamma}<<1$) and
high-ionization degree ($\tilde{\epsilon}>>1$), the dynamics of ions
can affect the motions of neutral particles so that they also land
in a larger radii to form the larger accretion disk as shown in
Fig.\ref{zRdiffmag}. The same behavior for increasing of accretion
disk radius via the effect of magnetic field can be deduced from
iso-density contours that are depicted in Fig.~\ref{isoden}.

\section*{Acknowledgments}
This work has been supported by grant of Research and Technology
Deputy of University of Mazandaran.


\clearpage

\clearpage

\clearpage

\clearpage

\clearpage

\end{document}